%% file: main.tex
\documentclass{article}
\usepackage{graphicx} 

\usepackage{arxiv}
\usepackage{dcolumn}
\usepackage{bm}
\usepackage[utf8]{inputenc}

\usepackage{natbib}
\usepackage{mathtools}
\usepackage{amsmath}
\usepackage{amssymb}
\usepackage{bm,bbm}
\usepackage{float}
\usepackage{color}
\usepackage{xcolor}
\usepackage{enumerate}

\usepackage{enumitem}

\usepackage{tikz}
\usepackage{comment}

\usepackage{url} 

\newcommand{\beq}{\begin{equation}}
\newcommand{\eeq}{\end{equation}}

\newcommand{\eps}{\varepsilon}

\input latex-defns.tex

\title{TURB-MHD: an open-access  database of forced homogeneous magnetohydrodynamic turbulence}

\author{
  Damiano Capocci \thanks{{Currently employed at the University of Edinburgh, School of Physics and Astronomy, United Kingdom.}} \\
  Dept. Physics and INFN \\
  University of Rome Tor Vergata, Italy.\\
  \texttt{dcapocci@ac.uk.ed}
\And
  Luca Biferale \\
  Dept. Physics and INFN\\
  University of Rome Tor Vergata and INFN, Italy\\
  \texttt{biferale@roma2.infn.it} \\
\And
  F. Bonaccorso \\
  Dept. Physics and INFN\\
  University of Rome Tor Vergata, Italy.\\
  \texttt{fabio.bonaccorso@roma2.infn.it} \\
\And
  Moritz Linkmann \\
  School of Mathematics and Maxwell Institute for \\
     Mathematical Sciences \\
  University of Edinburgh, UK \\
  \texttt{moritz.linkmann@ed.ac.uk} \\
}

\date{July 2024}

\begin{document}

\maketitle

\begin{abstract}
We present TURB-MHD, a database formed by six datasets of three-dimensional incompressible homogeneous magnetohydrodynamic turbulence maintained by a large-scale random forcing with minimal injection of cross helicity. Five of them describe a stationary state including one characterised by a \emph{weak} background magnetic field. The remaining dataset is non-stationary and is featured by a \emph{strong} background magnetic field. The aim is to provide datasets that clearly exhibit the phenomenon of the total energy cascade from the large to the small scales generated by the large-scale energy injection and one showing a partial inverse kinetic energy cascade from the small to the large scales. This database offers the possibility to realize a wide variety of analyses of fully developed magnetohydrodynamic turbulence from the sub-grid scale filtering up to the validation of an \emph{a posteriori} LES.\\
TURB-MHD is available for download using the SMART-Turb portal \url{http://smart-turb.roma2.infn.it}.
\end{abstract}

\section{Numerical simulations}\label{sec:numerics}
Data have been generated by direct numerical simulation of the 3D magnetohydrodynamic (MHD) equations
on a triply periodic domain of size $L_{\rm box} = 2\pi$ in each spatial direction.
Our case of interest is incompressible three-dimensional (3D)
homogeneous MHD turbulence. The datasets employ either standard-viscosity and hyperviscous dissipation law.
The dynamical variables are then the fluctuation velocity $ \vu (\vx, t)$
and the fluctuation magnetic field $ \vb (\vx,t) $, where we
measure the latter in Alfv\'en speed units:
        $  \bm{B} / \sqrt{4\pi \rho} \to \bm{B} $,
with $\rho$ the uniform mass density. In particular for the $z$ component, we set $\bm{B} = \bm{b} + B_0 \bm{\hat{e}_z}$ where $\bm{b}$ is the magnetic field fluctuations, $B_0$ the constant background magnetic field (BMF) and $e_z$ is $z$-axis unit vector.
The governing equations, with allowance for hyper-dissipation,
are:
\begin{align}
    &\frac{\partial\bm{u}}{\partial t} + \bm{u} \cdot \nabla \bm{u} = -\nabla \left(p + \frac{B^2}{2} \right) + \bm{B} \cdot \nabla \bm{B} + \nu_{\alpha} (-1)^{\alpha+1} \Delta^{\alpha} \bm{u} + \bm{F} \label{eq:mom_eq} \\
    &\frac{\partial \bm{B}}{\partial t} + \bm{u} \cdot \nabla \bm{B} =  \bm{B} \cdot \nabla \bm{u} + \mu_{\alpha} (-1)^{\alpha+1} \Delta^{\alpha} \bm{B} \label{eq:ind_eq} \\
    &\nabla \cdot \bm{u} =   0 \label{eq:divzeros_vel} \\
    &\nabla \cdot \bm{B} =   0  \label{eq:divzeros}
\end{align}
Here $p$ is the pressure divided by 
the fluid density which is constant,
 $ \nu_\alpha $ and $ \mu_\alpha $ are the hyper-viscosity and hyper-resistivity,
and
 $ \alpha $  denotes the power of the Laplacian operator employed in
 the hyper-dissipation.
Standard Laplacian dissipation corresponds to the case $\alpha =1 $. The forcing $\vF$ applied to the system is
a drag-free Ornstein--Uhlenbeck process, active in the wavenumber band $k \in [2.5,5.0] $ for each MHD simulation.
We consider both standard diffusive ($\alpha = 1$) and hyper-diffusive ($\alpha = 5$) cases, always with $ \nu_\alpha = \mu_\alpha $ \citep{borue1995forced}. \\

Equations \eqref{eq:mom_eq}-\eqref{eq:divzeros} are stepped forwards in time using a second-order Runge-Kutta scheme in a triply periodic $ (2\pi)^3 $ domain. Both the (hyper)viscous and magnetic (hyper)diffusion terms are treated implicitly using an integrating factor. The spatial discretisation was implemented via the standard pseudospectral method with complete dealiasing by the two-thirds rule \citep{PattersonOrszag71,CanutoEA}. 

Further details and mean values of key observables are summarised in table \ref{tab:datasets}. The hereby presented datasets have been used in \cite{damiano2024a} and \cite{damiano2024thesis} and they are being analysed for future works \citep{damiano2024b, damiano2024c, moritz2024, yasminthesis}.

\begin{table}
  \begin{center}
\def~{\hphantom{0}}
   \begin{tabular}{ccccccccccccccccc}
        \hline
        \hline
		 id & $\frac{B_0}{B_{\rm rms}}$ & $N$ &$\alpha$  & $E_u$ & $E_b$ & $\nu_\alpha$ & $\eps_u$ & $\eps_b$ & $L_u$ & $\tau$  & $\mbox{Re}$ & $\dfrac{k_\text{max}}{k_\text{diss}^u}$ & $\dfrac{k_\text{max}}{k_\text{diss}^b}$ & $\Delta t / \tau$ & \# \\
        \hline        
      A1 & 0 & 1024 & 1 & 0.75 & 0.33 & $4.2\times 10^{-4}$ & 0.22  & 0.53 & 0.55  & 0.77  & 936  & 1.46 & 1.17 & 1.3 & 11  \\
      A2 & 0  & 2048 & 1 & 0.73 & 0.38  & $2.0\times 10^{-4}$ & 0.22  & 0.52 & 0.55  & 0.79  & 2144  & 1.68   & 1.35 &  1.1 & 18 \\
      A3 & 0  &  1024 & 5 & 0.70 & 0.48 & $5 \times 10^{-23}$ & 0.31 & 0.43 & 0.53  & 0.78  & 4272  & 1.45 & 1.43 & 1.5  & 26  \\
      A4 & 0 &  2048 & 5 & 0.66  & 0.54  & $5 \times 10^{-26}$ & 0.33  & 0.43  & 0.51  & 0.81    & 9931 & 1.38 & 1.37 & 1.1  &  18  \\
       \hline
      C1 &  1.2 & 1024 & 5 & 0.73 & 0.76 & $5 \times 10^{-23}$ & 0.32 & 0.42 & 0.44  & 1.05   & 2742   & 1.45 & 1.43 & 1.1  & 26  \\
      C10 &  12.7 & 1024 & 5 & 3.52 & 0.31 & $5 \times 10^{-23}$ & 0.32 & 0.40 & 1.56  & 1.08   & 7501   & 1.45 & 1.44 & 0.8  & 19  \\

        \hline
      \hline
        \end{tabular}
        \caption{Simulation parameters and key observables, where
        $N$ is the number of collocation points in each coordinate,
        $\alpha$ is the power of $\nabla^2$ used in the hyper-diffusion,
        $E_u$ the mean total kinetic energy,
        $\nu_\alpha$ the kinematic (hyper)viscosity,
        $\eps_u$ and $\eps_b$ are the kinetic and Joule energy dissipation rates,
        $L_u = (3 \pi/4 E_u) \int_0^{k_\text{max}} dk \ E_u(k)/k$ the
        integral scale,
        $\tau = L_u/\sqrt{2E_u/3}$ the large-scale eddy-turnover time, $B_{\rm rms} = \sqrt{2E_b}$ the root-mean-square value of the magnetic field fluctuations,  and
        $\mbox{Re}$ is the (integral scale) Reynolds number. 
	  Moreover, $k_\text{diss}^u$ and $k_\text{diss}^b$ are the wavenumbers associated with the hyperdiffusive Kolmogorov scales $\eta^u_{\alpha} = (\nu_\alpha^3 / \varepsilon_u)^{1 / (6  \alpha -2)}$ and $\eta^b_{\alpha} = (\mu_\alpha^3 / \varepsilon_b)^{1 / (6  \alpha -2)}$ respectively calculated in terms of the viscous and Joule dissipation rates,
        $k_\text{max}$ the largest retained wavenumber component after dealiasing, 
        $\Delta t$ the mean of the snapshots sampling intervals,
        and \# indicates the number of snapshots used in the statistics.
        The magnetic Prandtl number,  
        $Pm = \nu_\alpha / \mu_\alpha $ defines the ratio between the hyperviscosity and magnetic hyperdiffusivity, equals unity for each dataset. Regarding C10, the parameters $E_u$, $E_b$, $L_u$, $\tau$ and $Re$ , are related to the last stationarity interval of the run corresponding to $t/\tau \in [95, 98] $ (see fig.~\ref{fig:B0_10}). 
        }
   \label{tab:datasets}
  \end{center}
\end{table}
The spatial resolution of the simulations can be quantified by both the
grid spacing $\Delta{x} = 2\pi / N$
and the (hyper-diffusive) Kolmogorov scales
        $ \eta^u_{\alpha} = (\nu_\alpha^3 / \varepsilon_u)^{1/(6  \alpha -2)} $, and 
        $ \eta^b_{\alpha} = (\mu_\alpha^3 / \varepsilon_b)^{1/(6  \alpha -2)} $,  
where $\varepsilon_u$ and $\varepsilon_b$ are the mean kinetic and magnetic energy dissipation rates.
For adequate resolution we require $ \eta^u_\alpha / \Delta{x} \gtrsim 1.3 $
and $ \eta^b_\alpha / \Delta{x} \gtrsim 1.3 $
  \citep[e.g.,][]{DonzisEA08, WanEA10-accuracy}.

\begin{figure}
\vspace{-0.8cm}
	\begin{center}
         \noindent\makebox[\textwidth]{
         \includegraphics[width=.6\columnwidth]{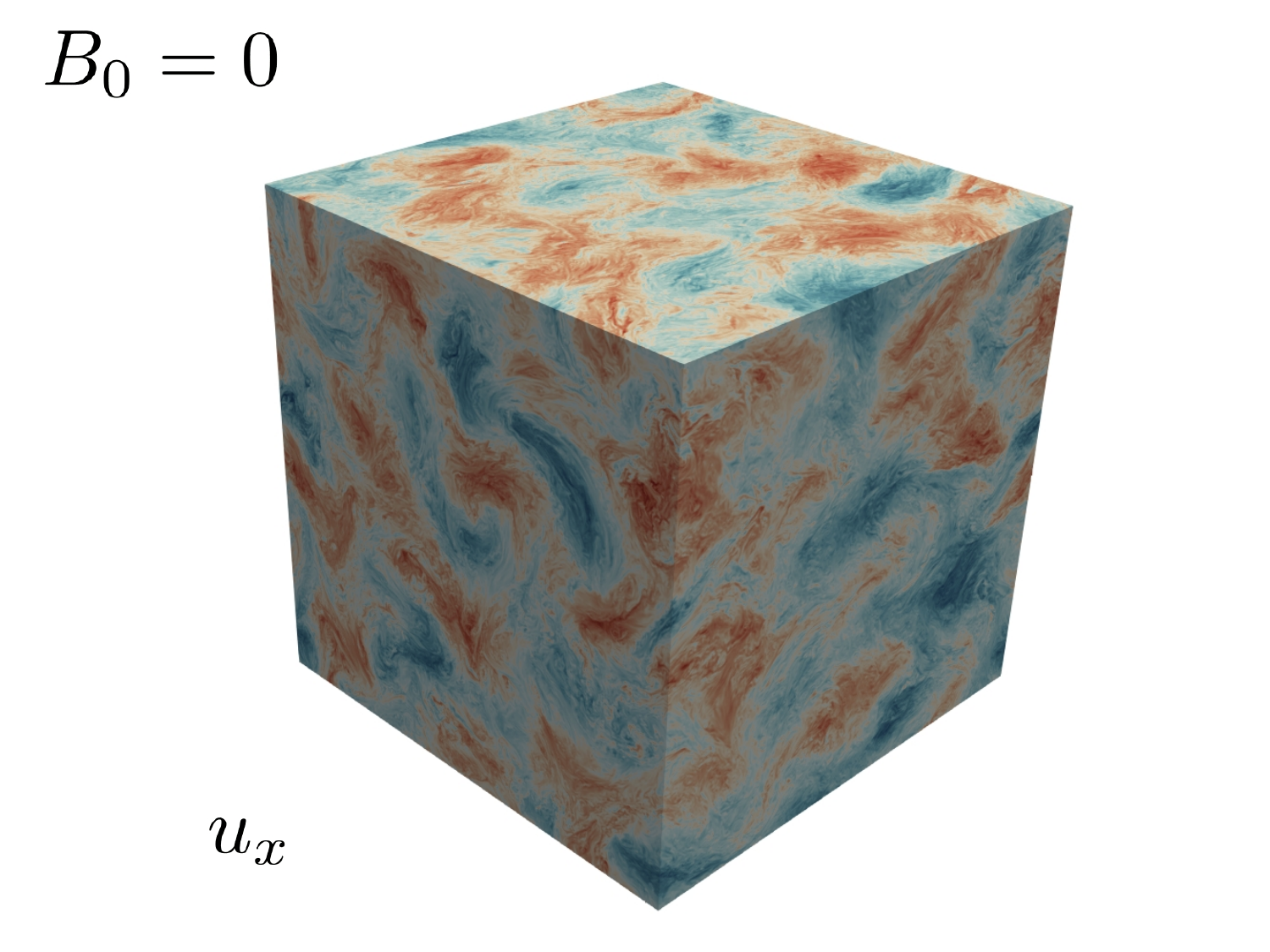}
         \hspace{-0.8cm}   
         \includegraphics[width=.6\columnwidth]{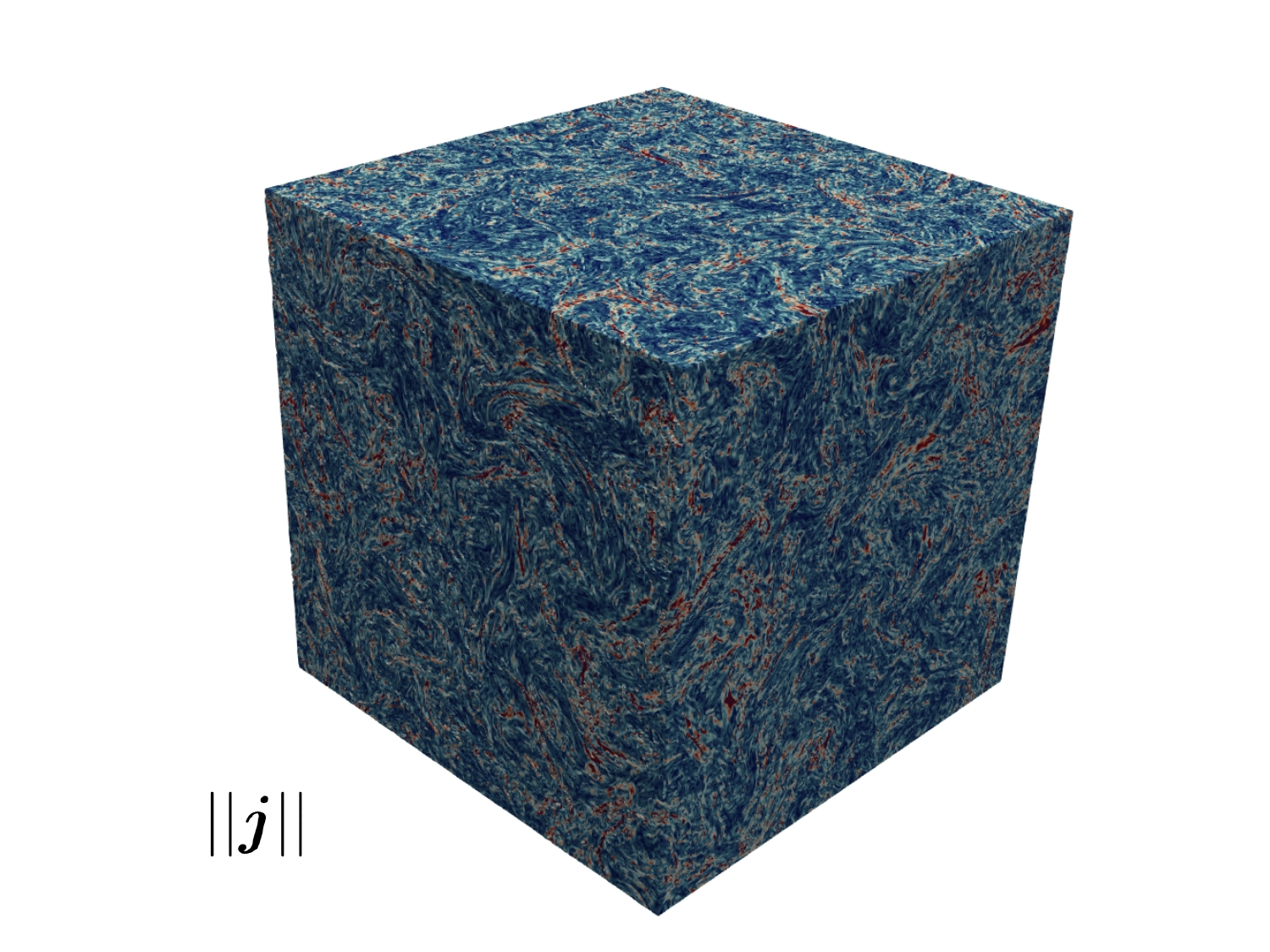}  
         }\vspace{0.7cm}
         \noindent\makebox[\textwidth]{
         \includegraphics[width=.6\columnwidth]{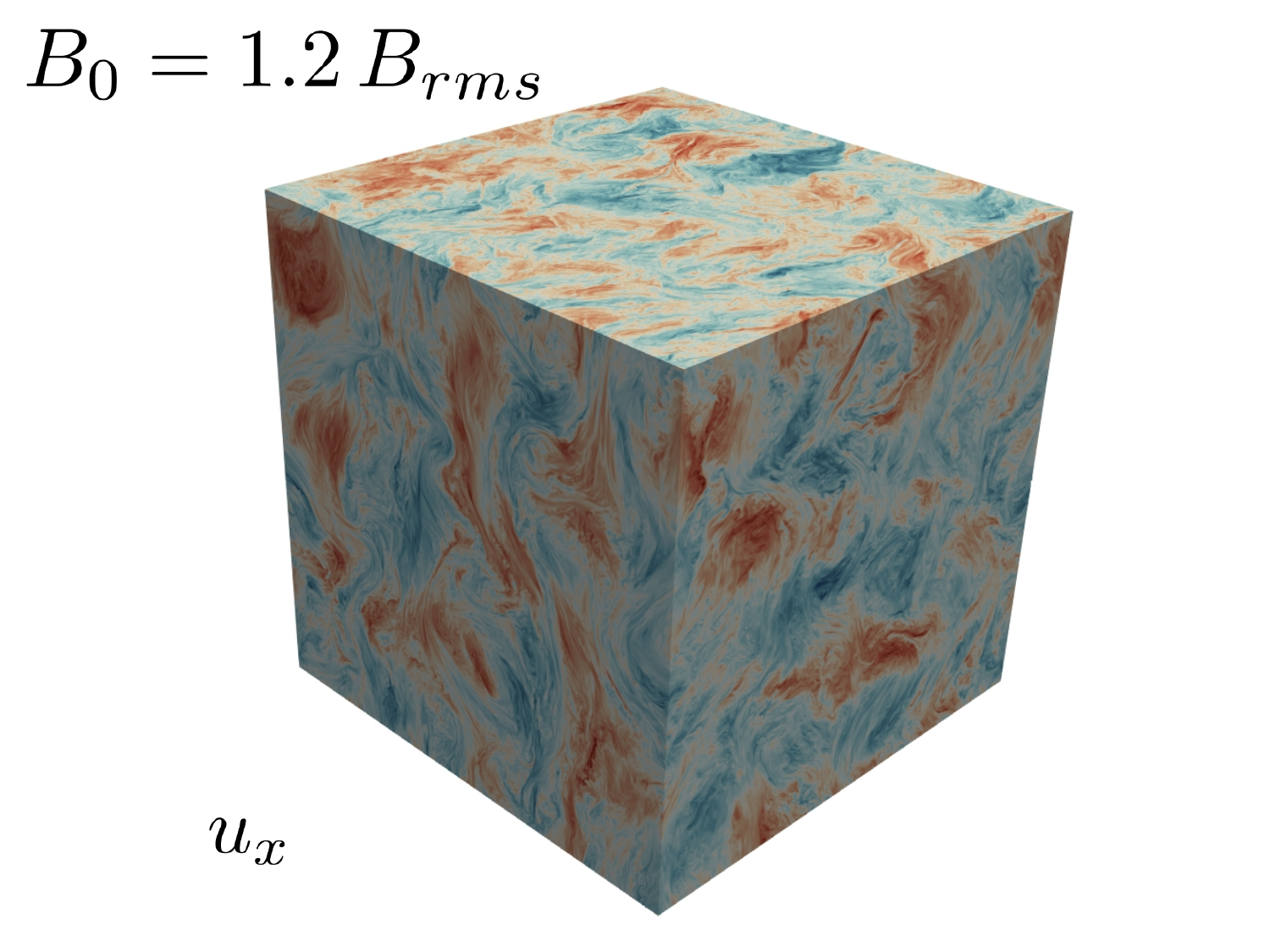} 
         \hspace{-0.8cm}   
         \includegraphics[width=.6\columnwidth]{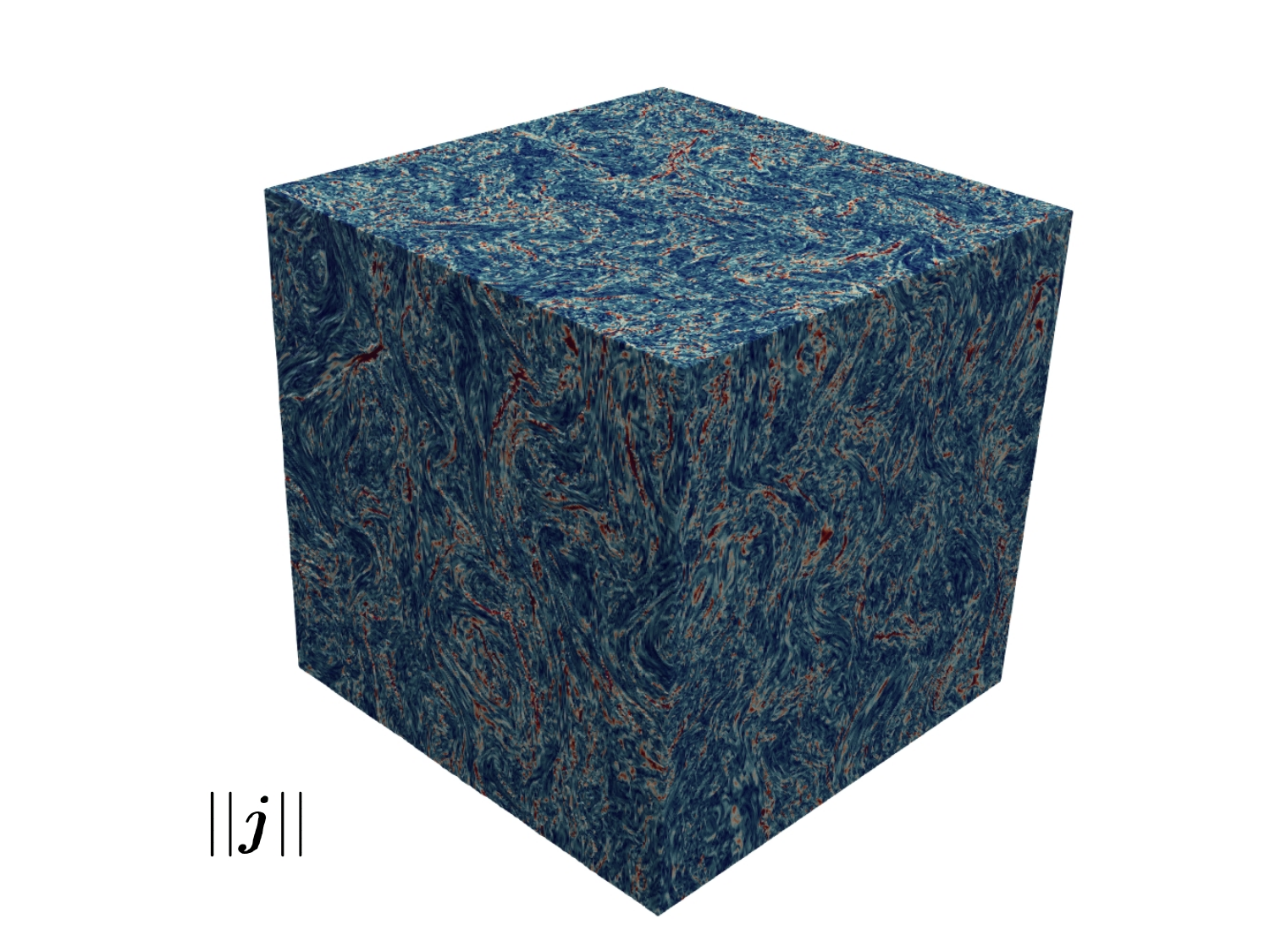} 
         }\vspace{0.7cm}
         \noindent\makebox[\textwidth]{
         \includegraphics[width=.6\columnwidth]{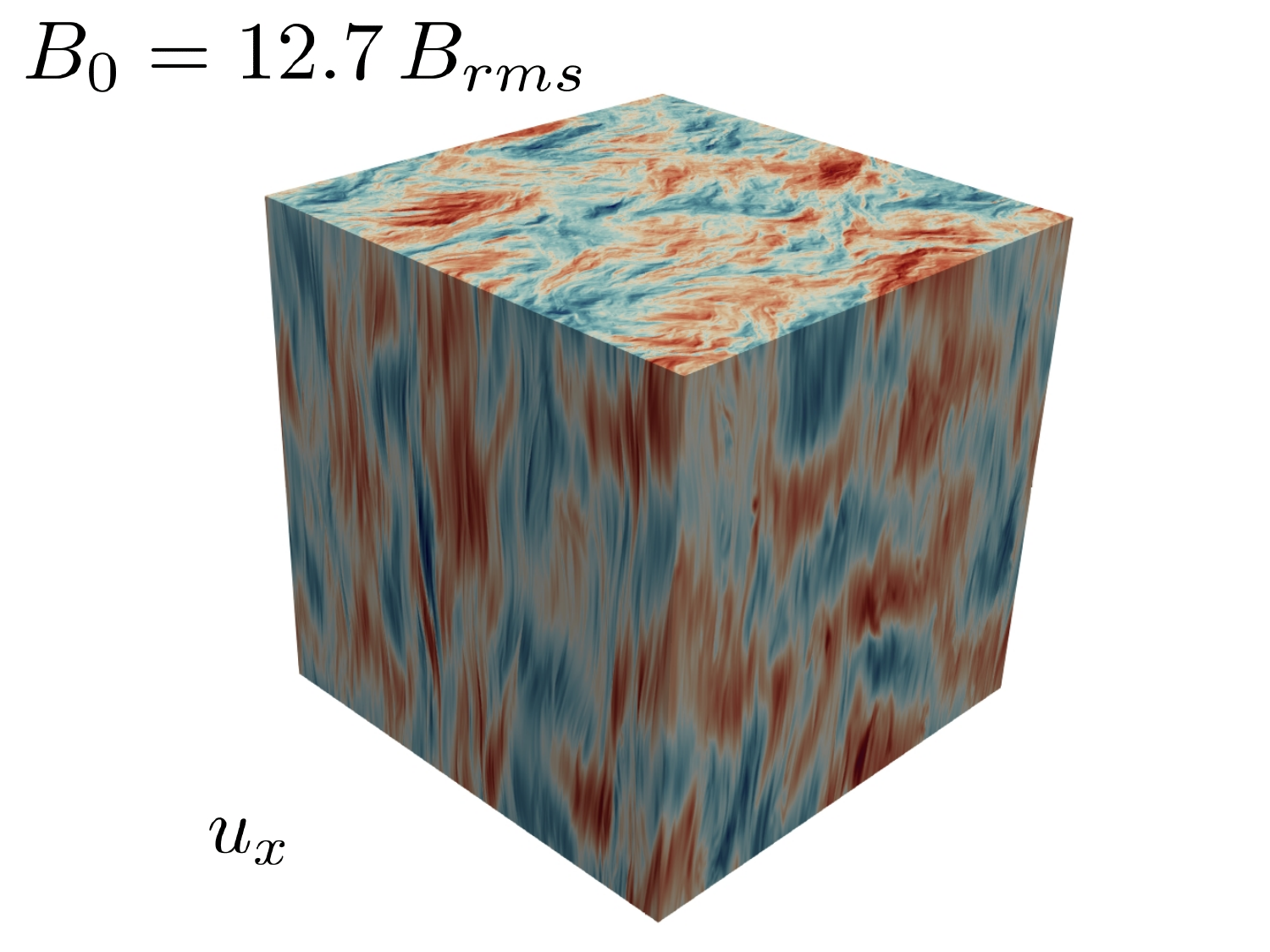} 
         \hspace{-0.8cm}   
         \includegraphics[width=.6\columnwidth]{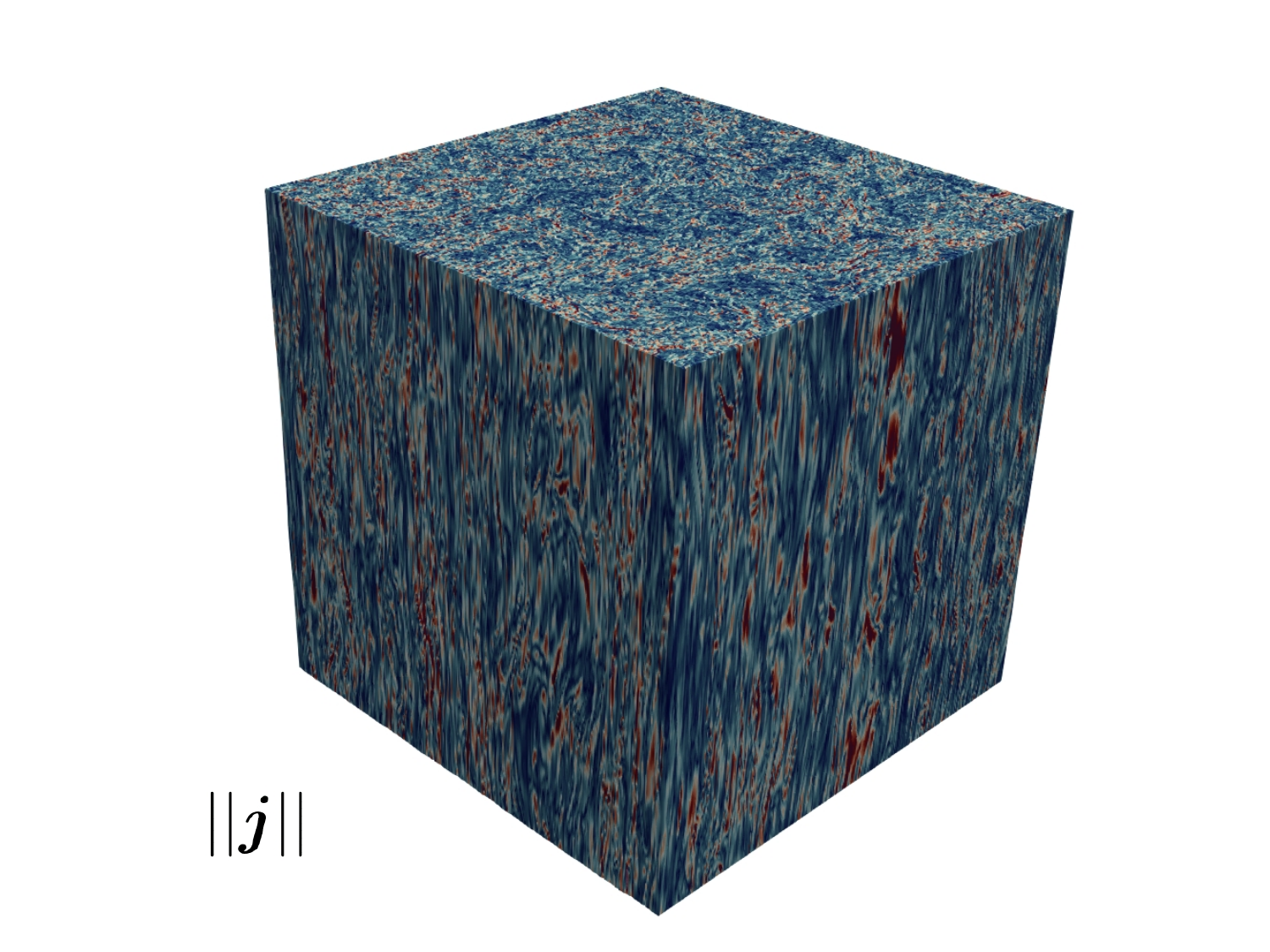} 
         }\vspace{-0.3cm}
    \end{center}
	 \caption{3D visualization where each horizontal panel is formed by three visualization cubes showing $u_x$ and the magnitude of $\bm{j} = \nabla \times \bm{b}$ as a function of the position $(\bm{x},\bm{y}$, $\bm{z})$. Top panel $B_0=0$, middle panel $B_0=1.2 \, B_{rms}$ and bottom panel $B_0=12.7 \, B_{rms}$. In each visualisation the colour range is shared between the configuration. Although the colour ranges are the same, the bottom left panels shows a more intense colouring due to the higher kinetic energy, see table \ref{tab:datasets}.}
\label{fig:visuals}
\end{figure}

\begin{figure}
    \centering
    \includegraphics[width=1.0\columnwidth]{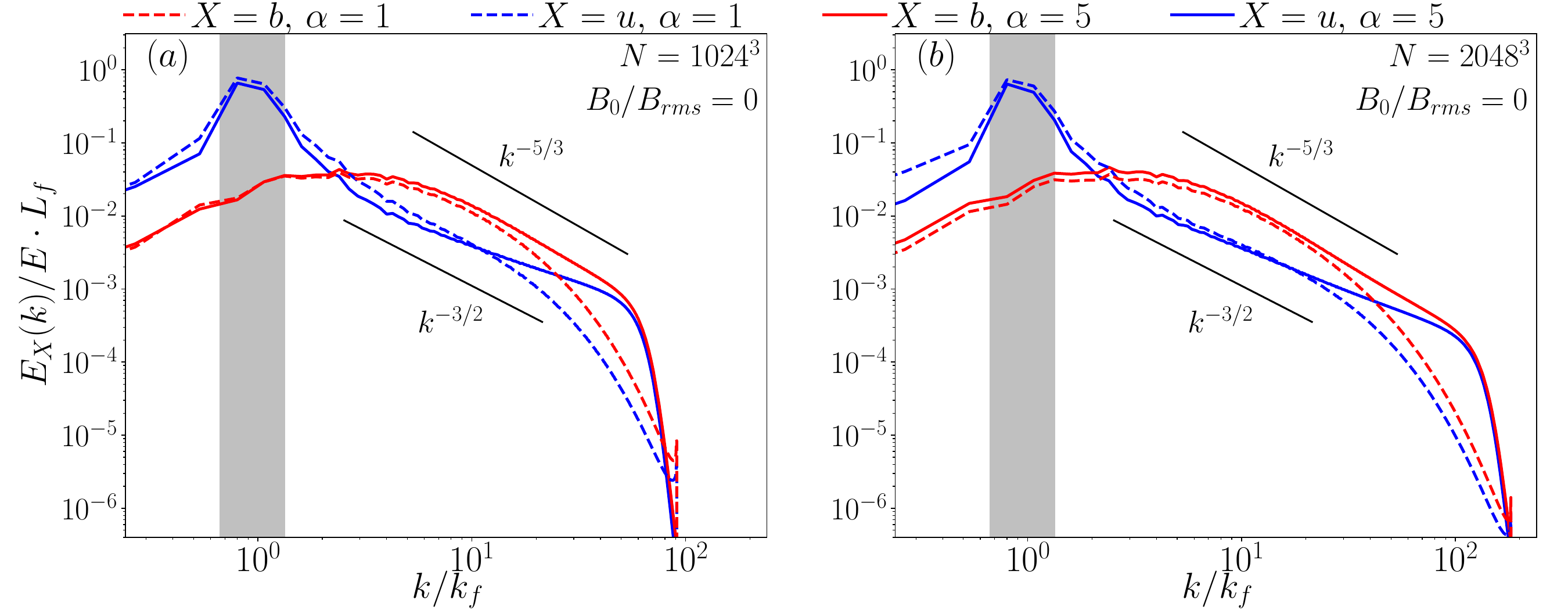}
    \includegraphics[width=.49\columnwidth]{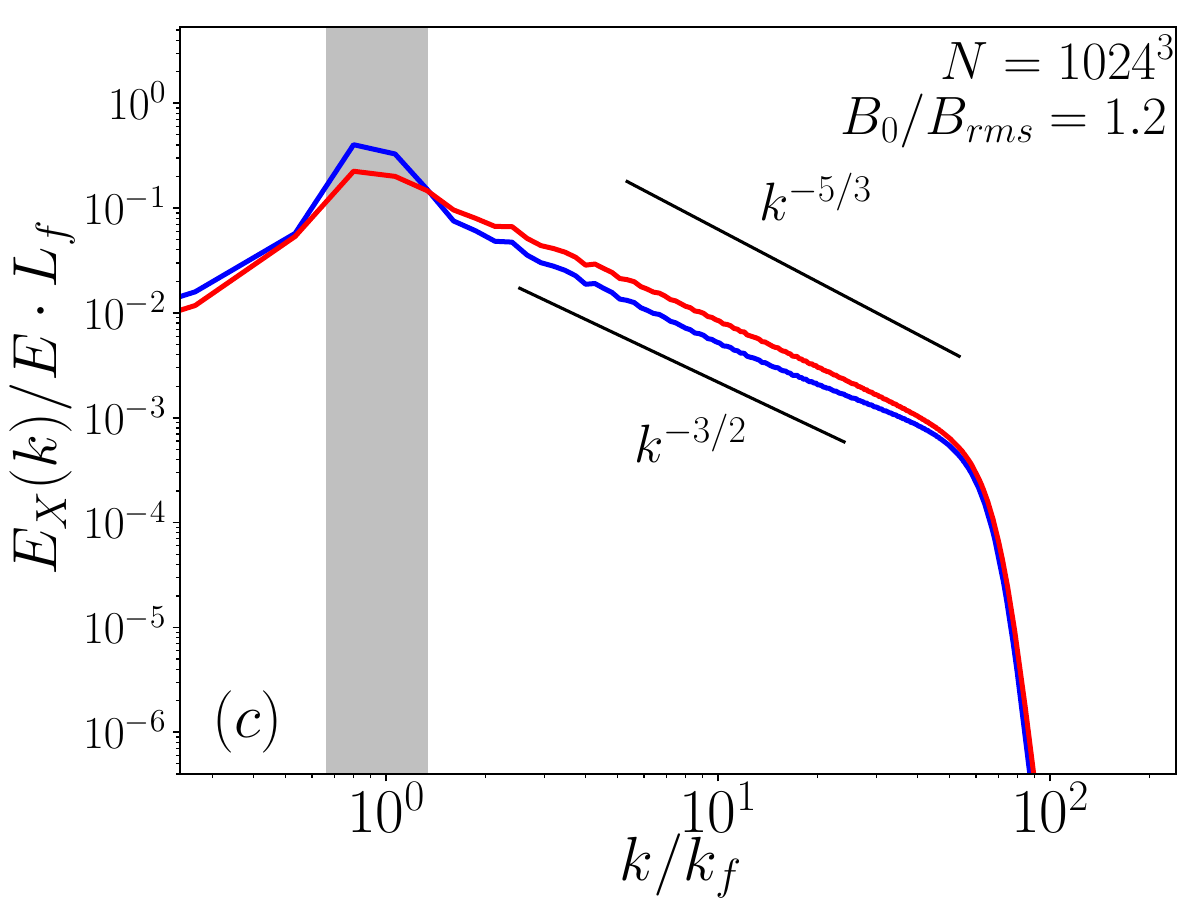}
    \includegraphics[width=.49\columnwidth]{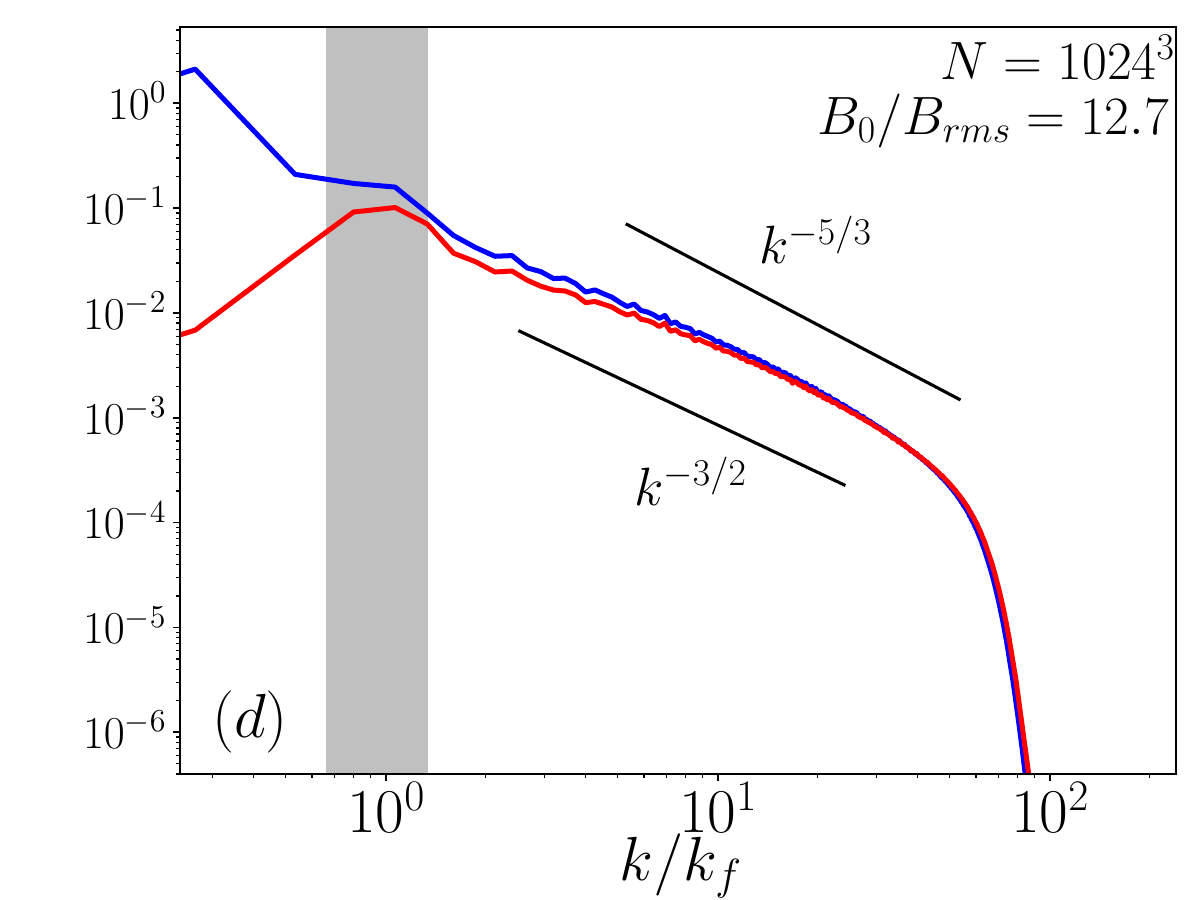}
    \caption{Time-averaged omnidirectional spectra for the velocity and magnetic field, respectively $E_u(k)$ and $E_b(k)$ normalised by the the product between the mean total energy and the forcing scale, both indicated in table \ref{tab:datasets}, as functions of the adimensional variable $k / k_f = L_f/\ell$: (a) comparison between datasets A1 and A3, (b) between datasets A2 and A4 while (c) and (d) refers to dataset C1 and C10 respectively. The gray region indicates the wavenumber band $k \in [2.5,5.0]$ where the velocity field is forced. }
    \label{fig:spectra_all}
\end{figure}

\begin{figure}
    \centering
    \includegraphics[width=0.49\columnwidth]{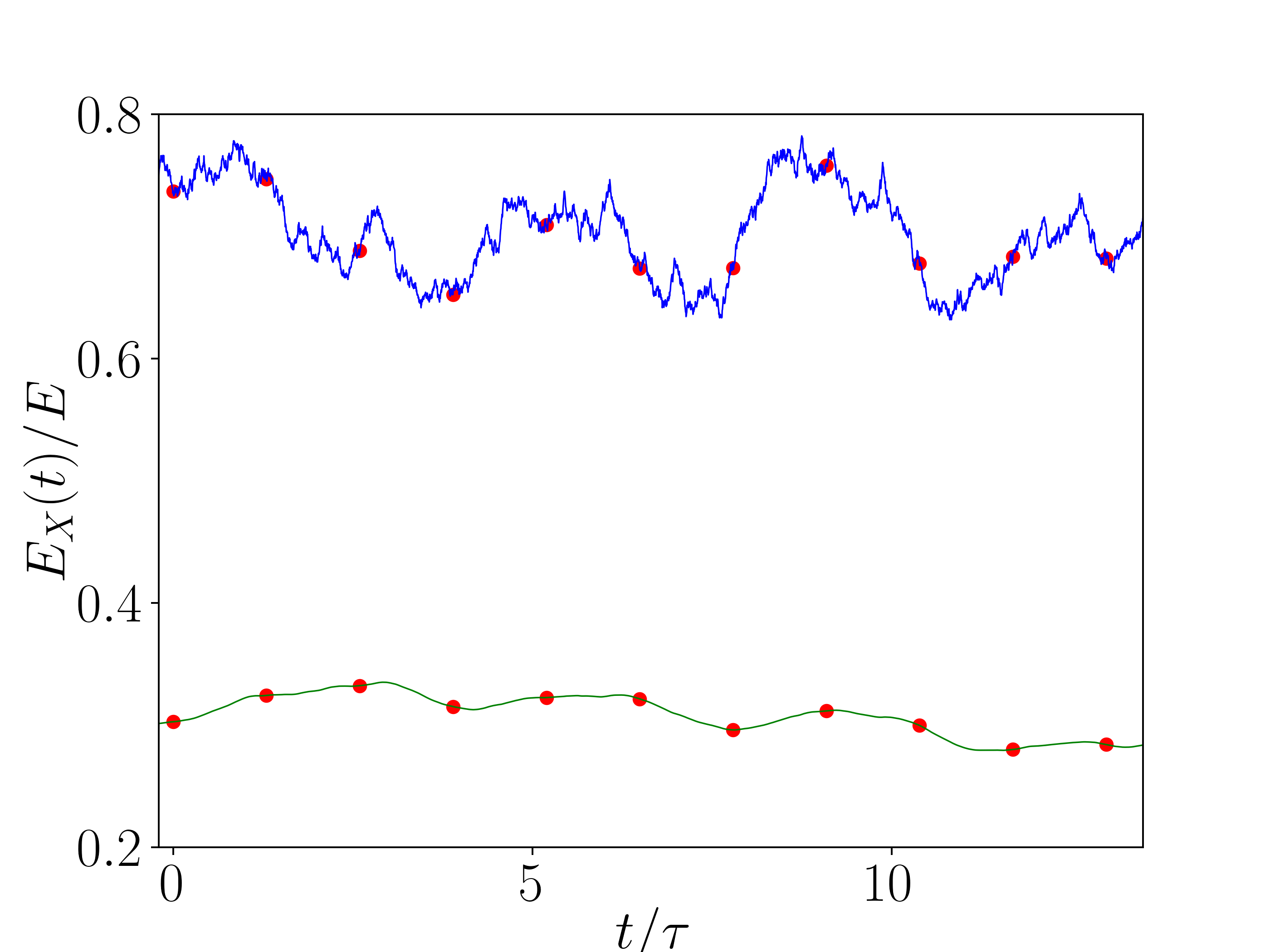}
    \includegraphics[width=0.49\columnwidth]{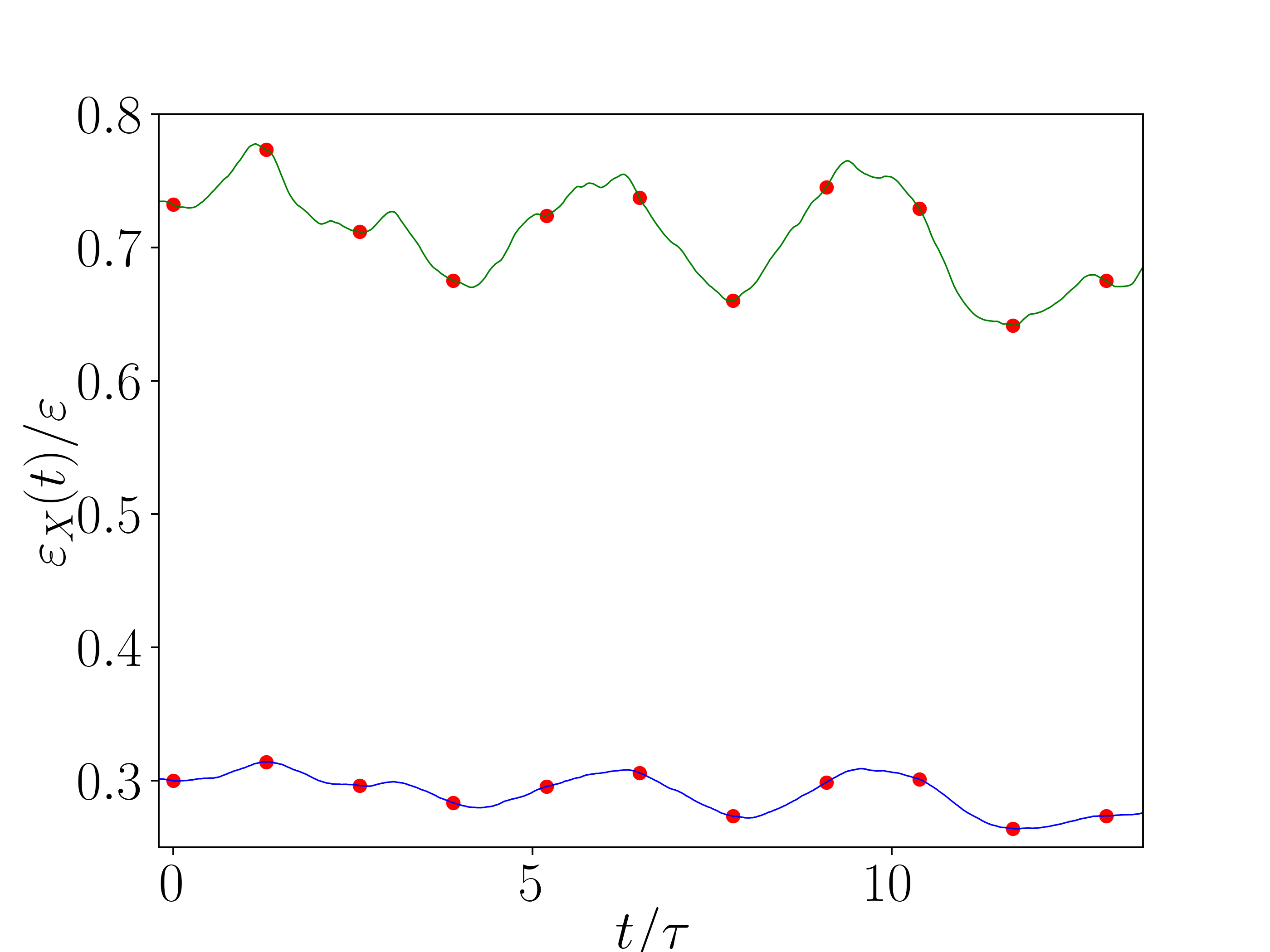}

    \includegraphics[width=0.49\columnwidth]{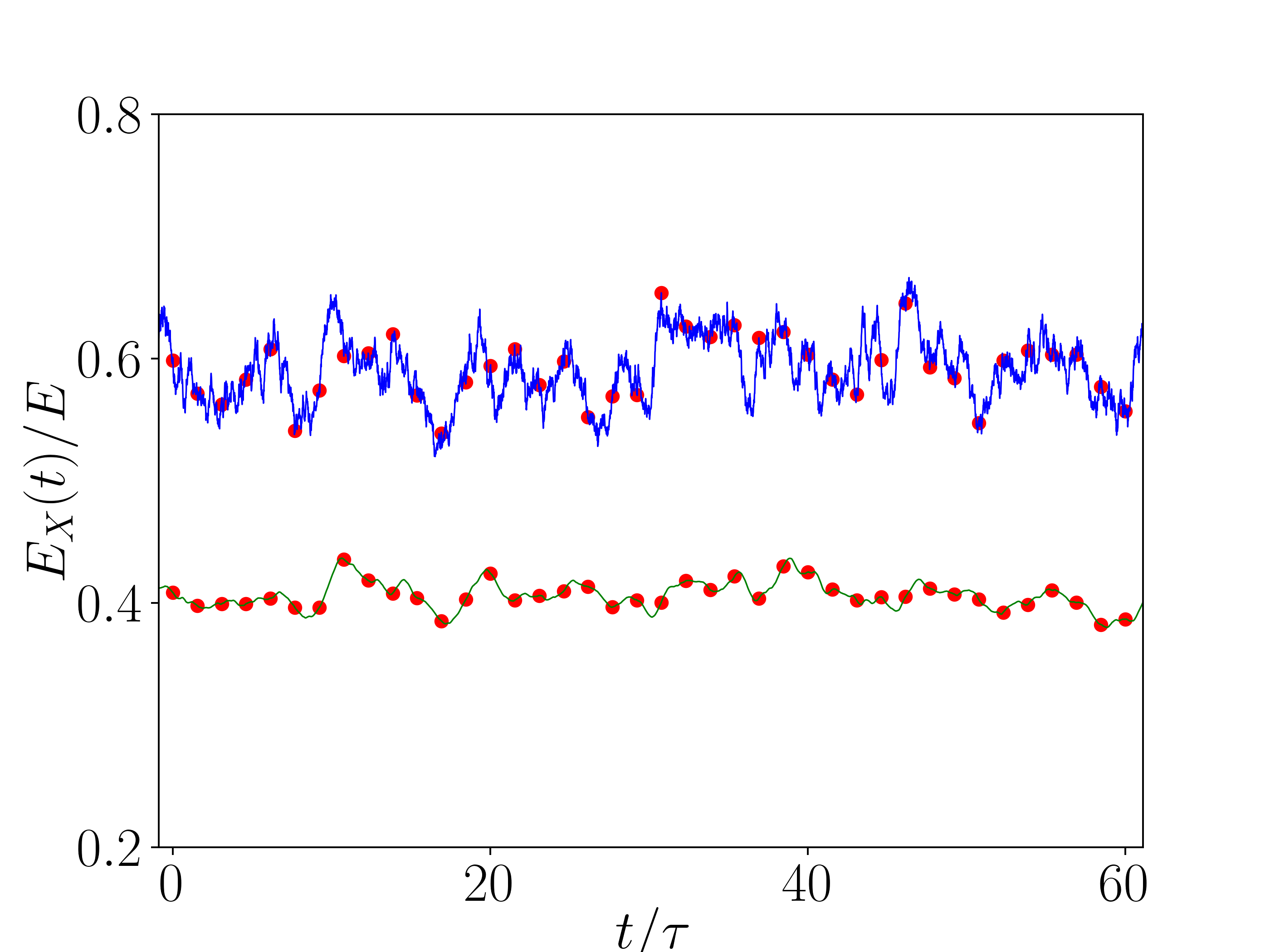}
    \includegraphics[width=0.49\columnwidth]{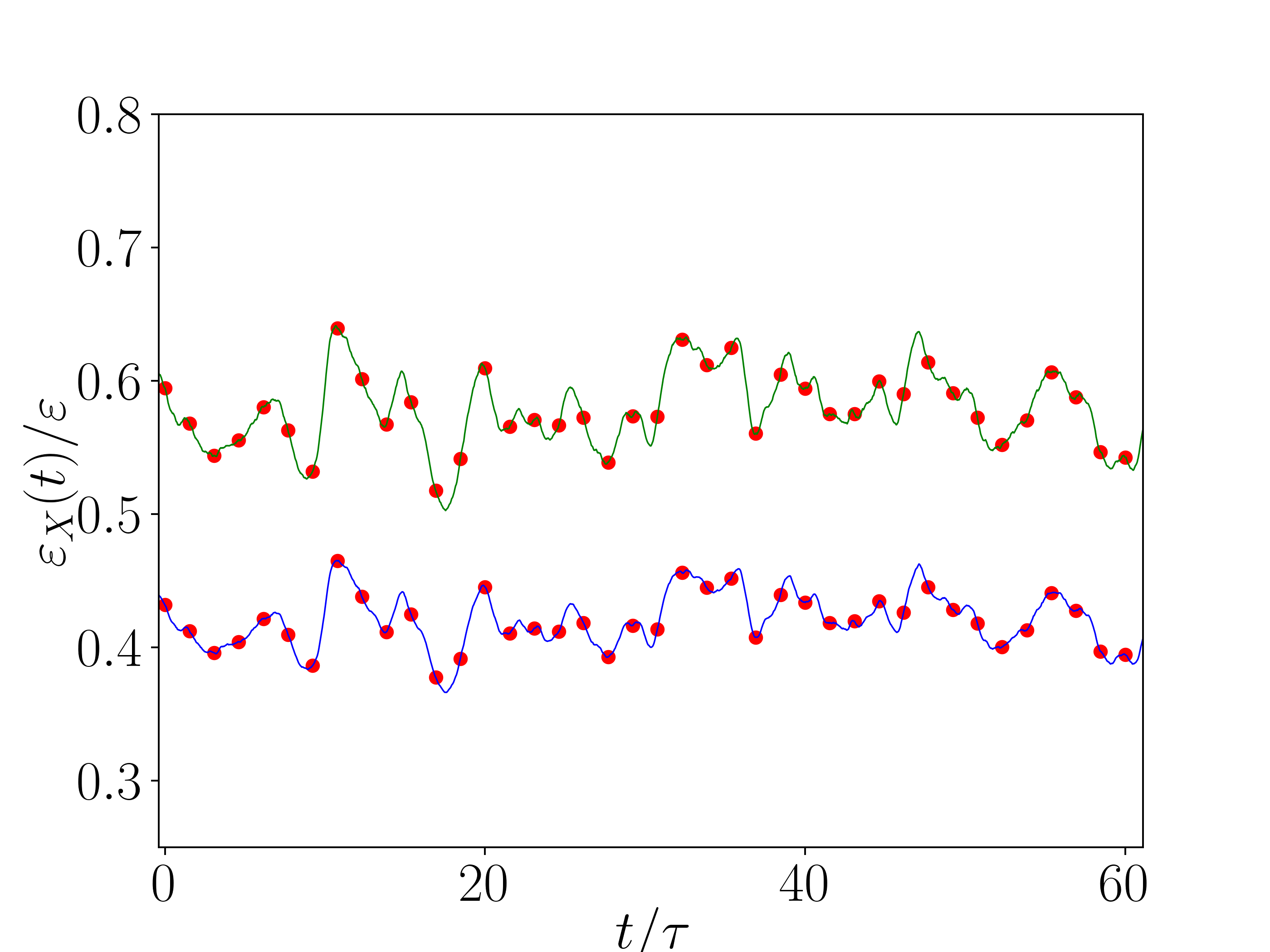}
    \caption{Time evolution of global observables for datasets A1 and A3. Left panel: Time evolution of mean kinetic and magnetic energy normalised by the mean total energy. Right panel: mean kinetic and Joule dissipation rate normalised by the total energy dissipation. The red dots correspond to the sampled velocity and magnetic field configurations. }
    \label{fig:hyperv_1024}
\end{figure}

\begin{figure}
    \centering
    \includegraphics[width=0.49\columnwidth]{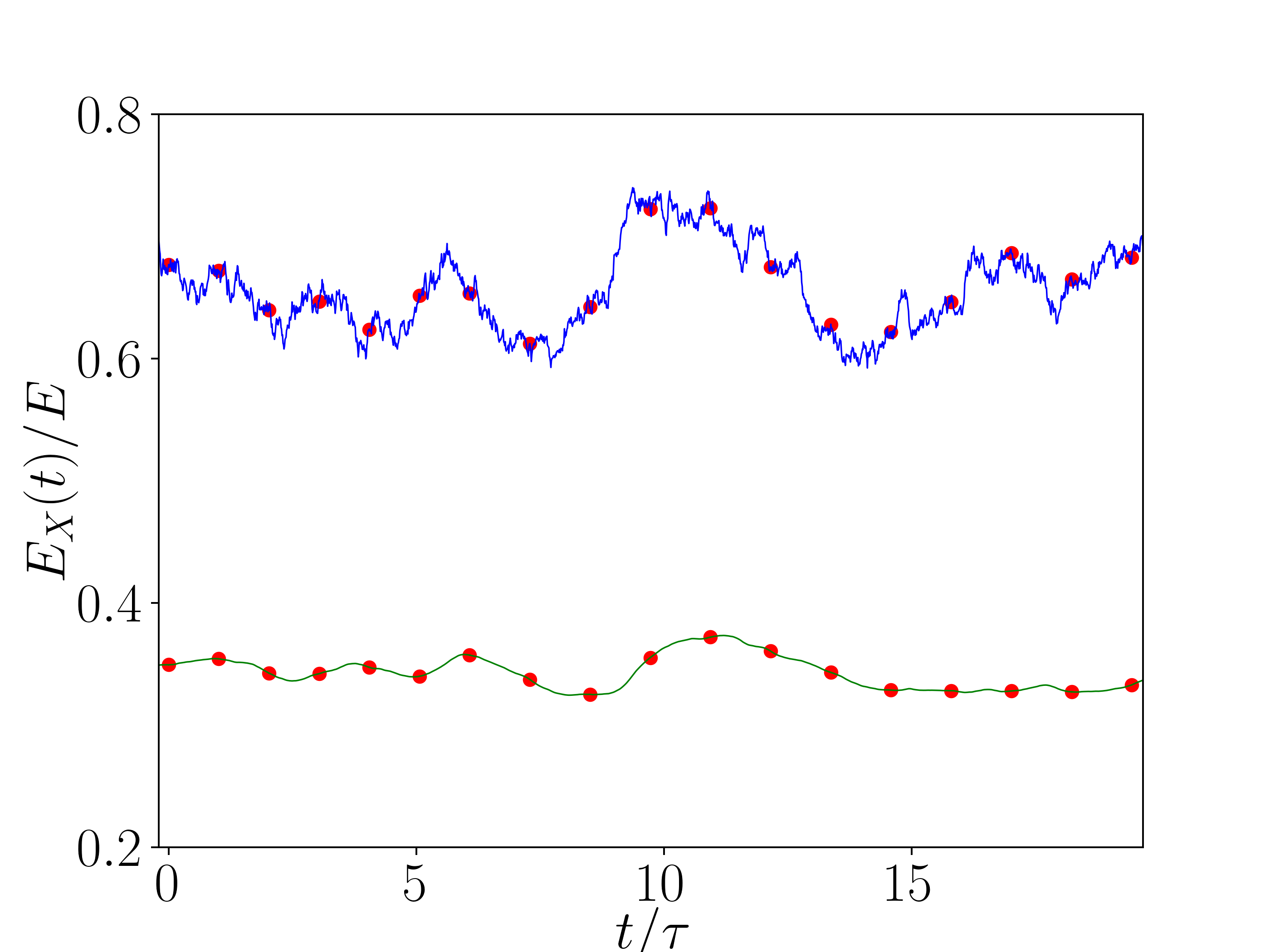}
    \includegraphics[width=0.49\columnwidth]{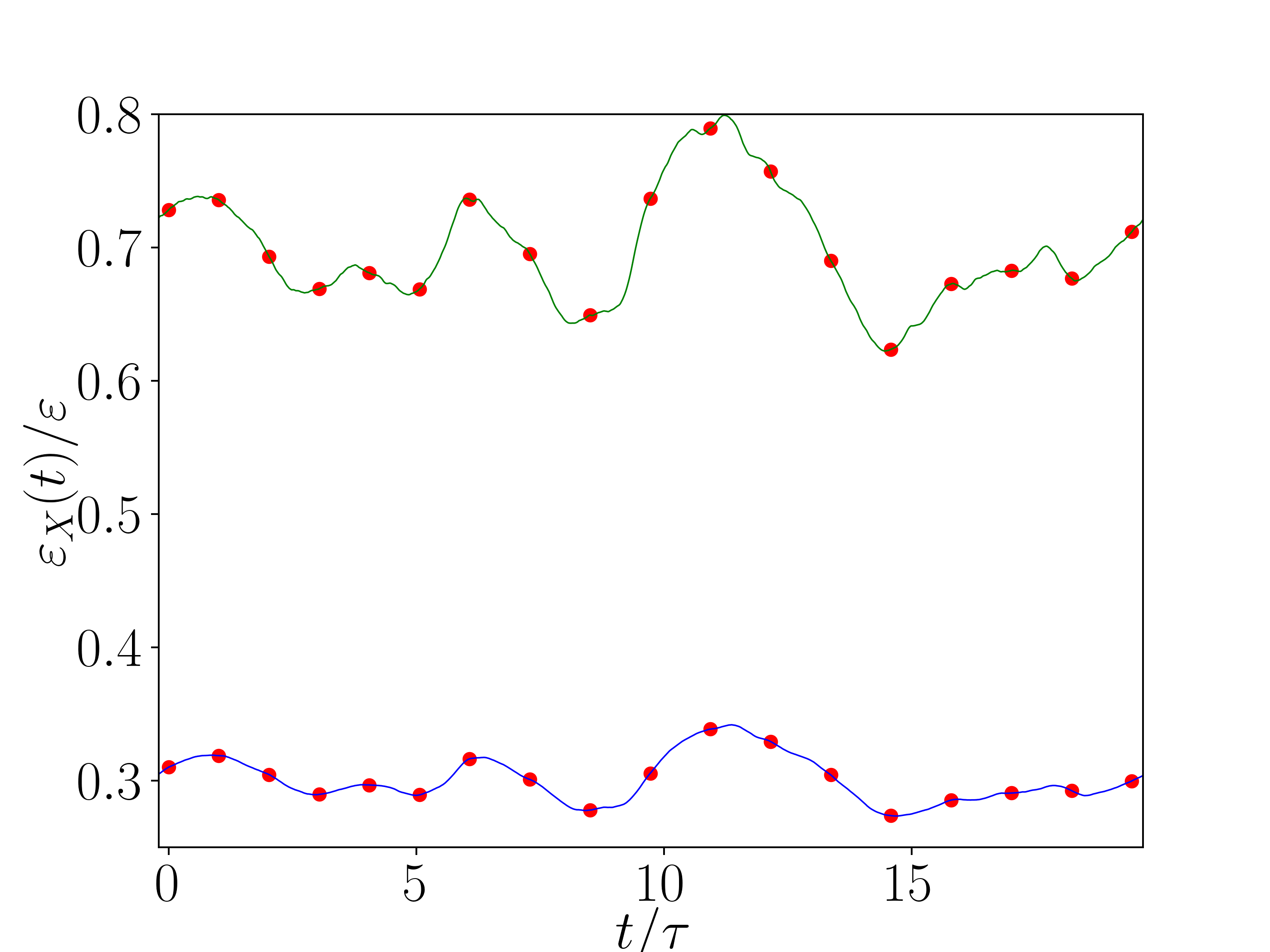}

    \includegraphics[width=0.49\columnwidth]{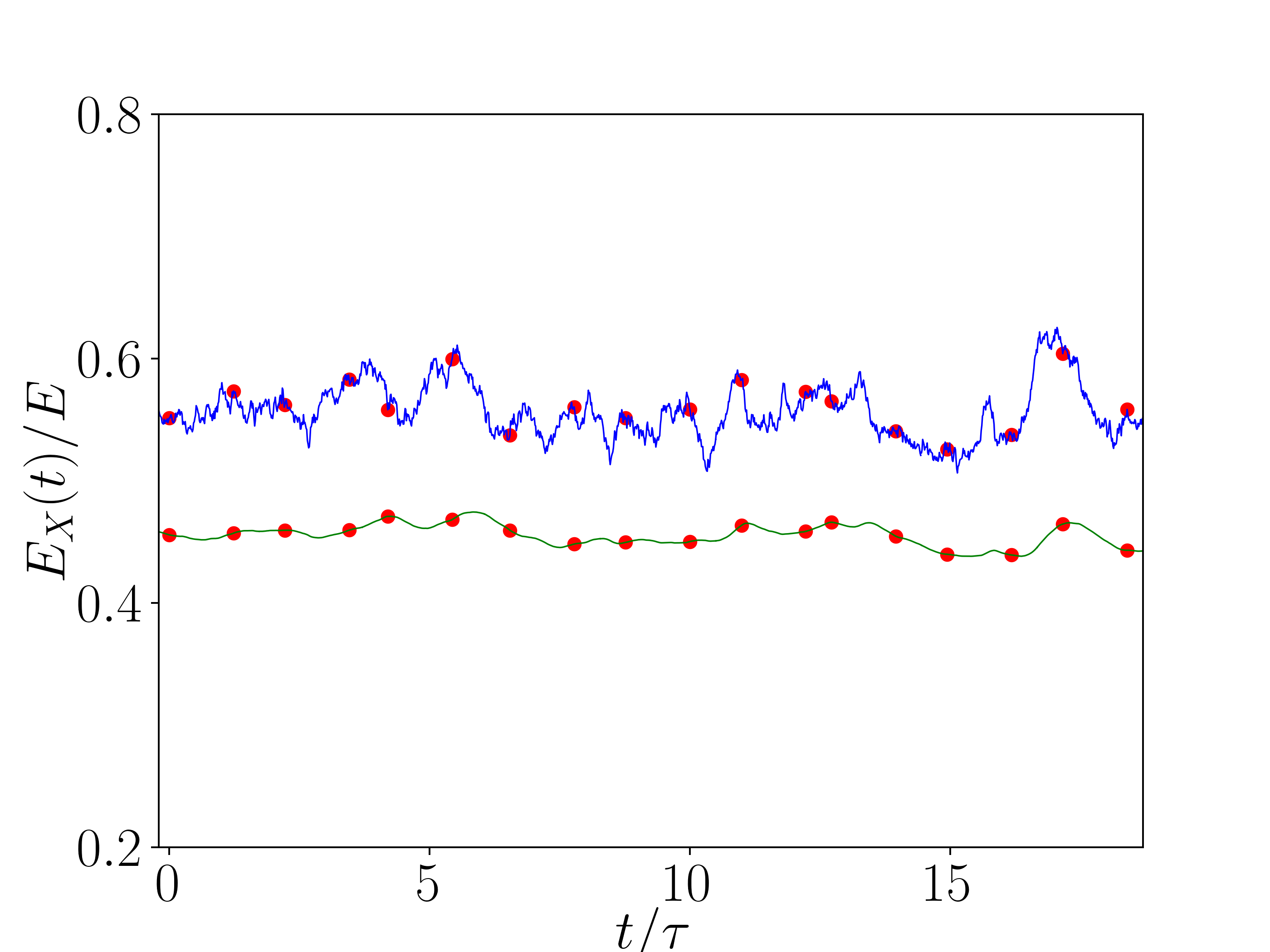}
    \includegraphics[width=0.49\columnwidth]{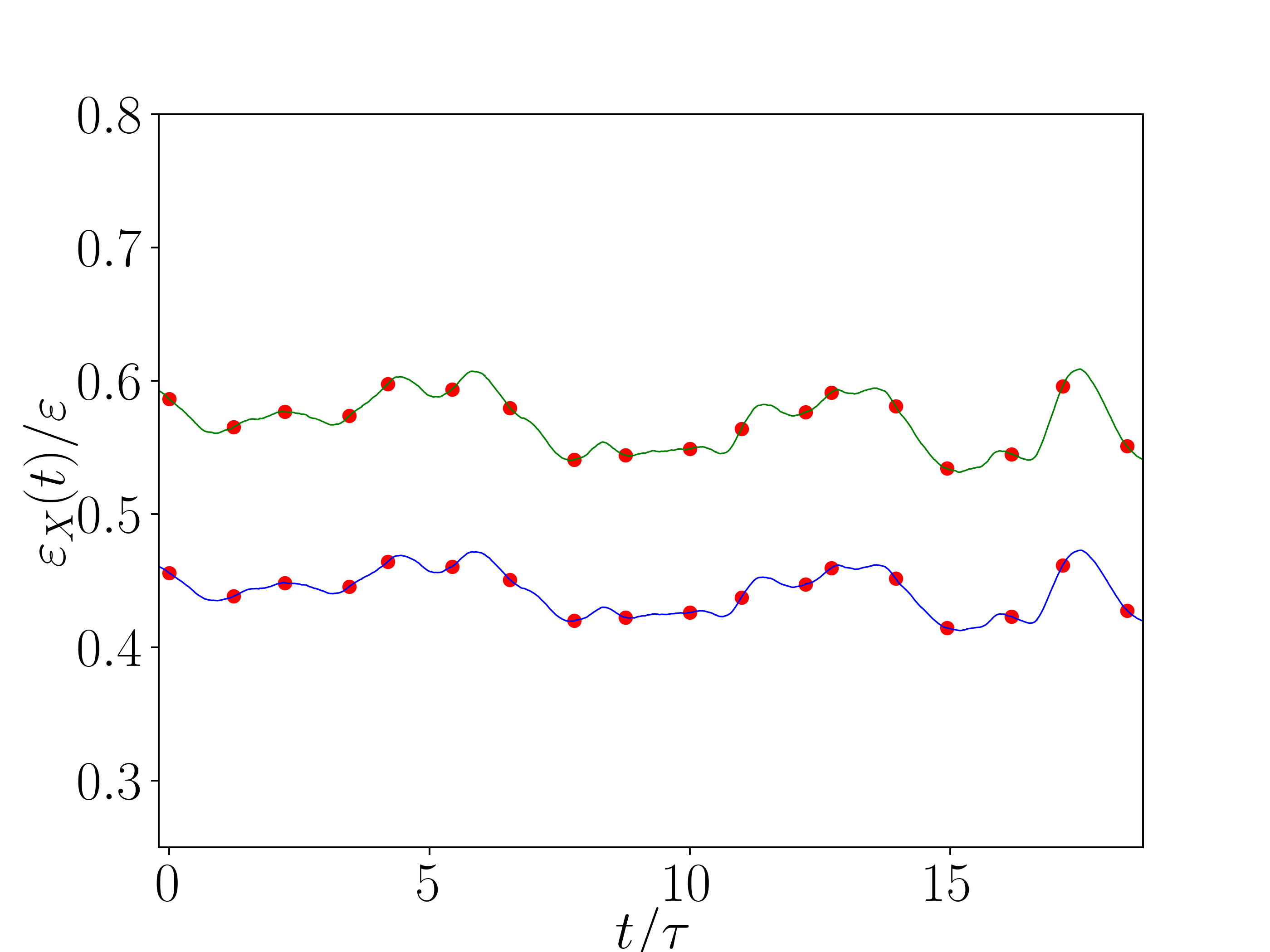}
    \caption{Time evolution of global observables for datasets A2 and A4. The y-axes ranges are the same as those in fig.~\ref{fig:hyperv_1024}. }
    \label{fig:hyperv_2048}
\end{figure}

\begin{figure}
    \centering
    \includegraphics[width=0.49\columnwidth]{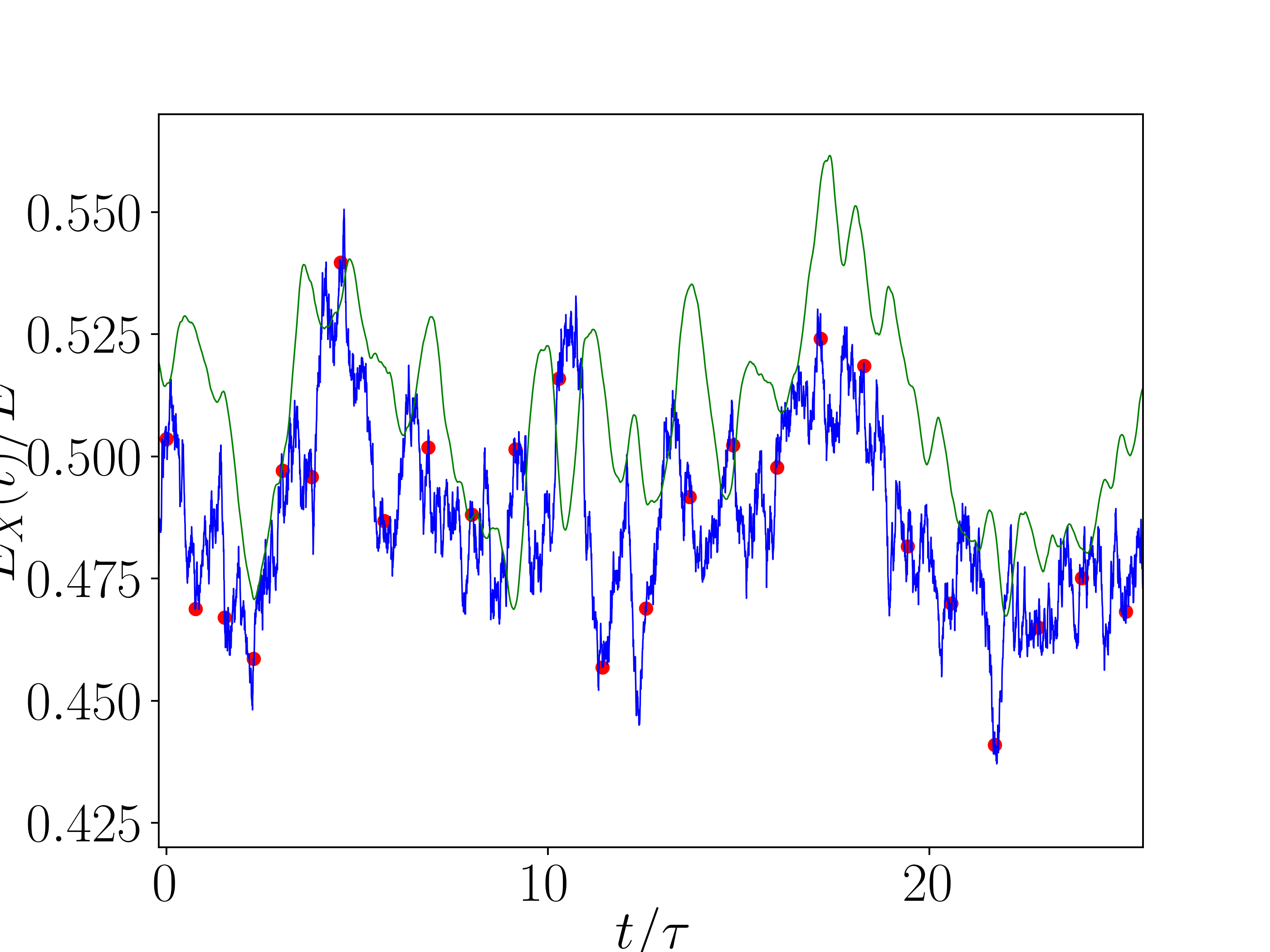}
    \includegraphics[width=0.49\columnwidth]{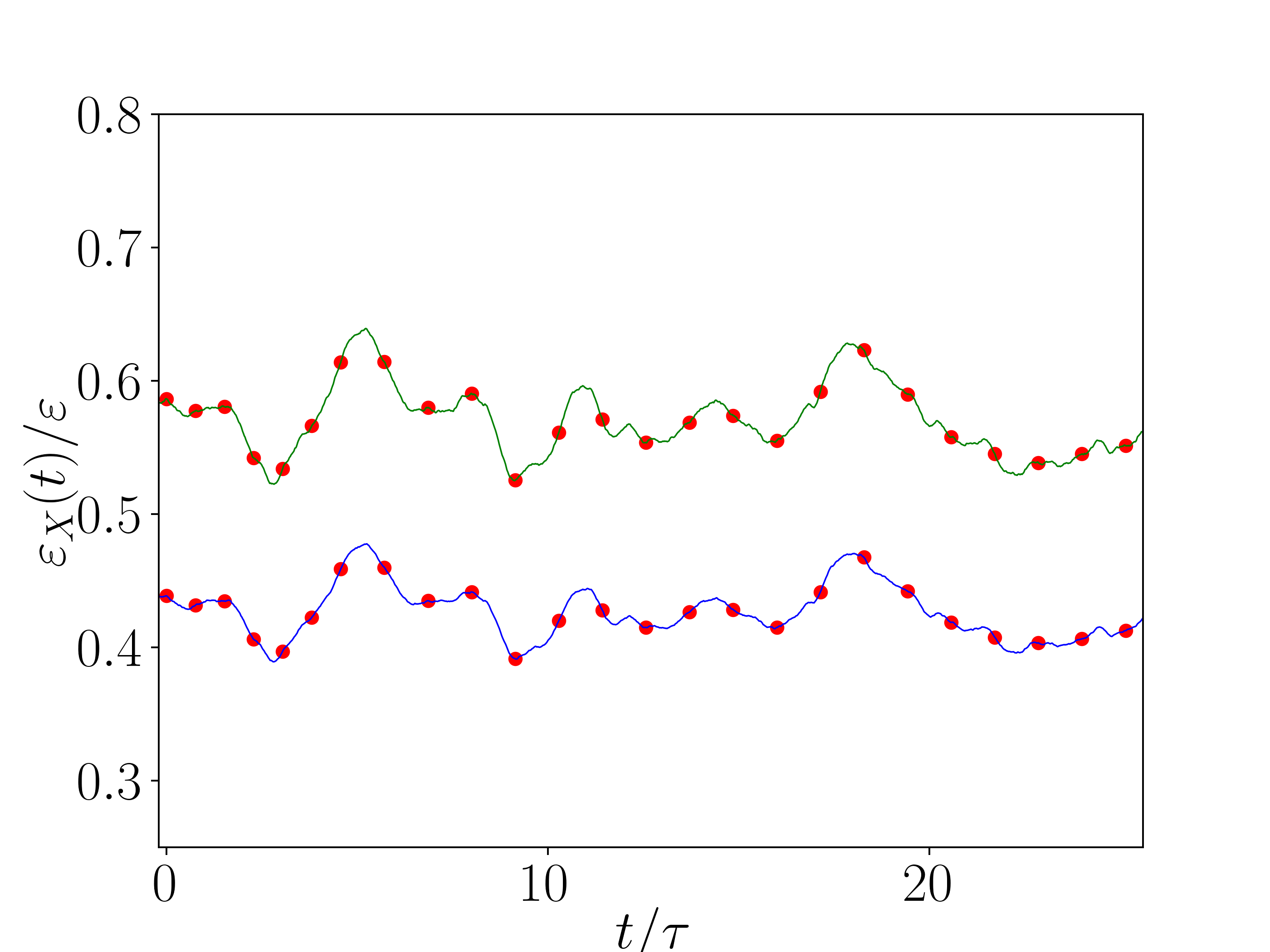}
    \caption{Analogously to fig.~\ref{fig:hyperv_1024}, time evolution of global observables for dataset C1. To improve the clarity of the picture, the red markers indicating the sampled configurations in the magnetic field have been omitted to provide better clarity of the timeseries. In addition, for clarity sake, the y-axis range of the left panel is narrower than that of the corresponding left panels above.}
    \label{fig:B0_1}
\end{figure}

\begin{figure}
    \centering
    \includegraphics[width=0.49\columnwidth]{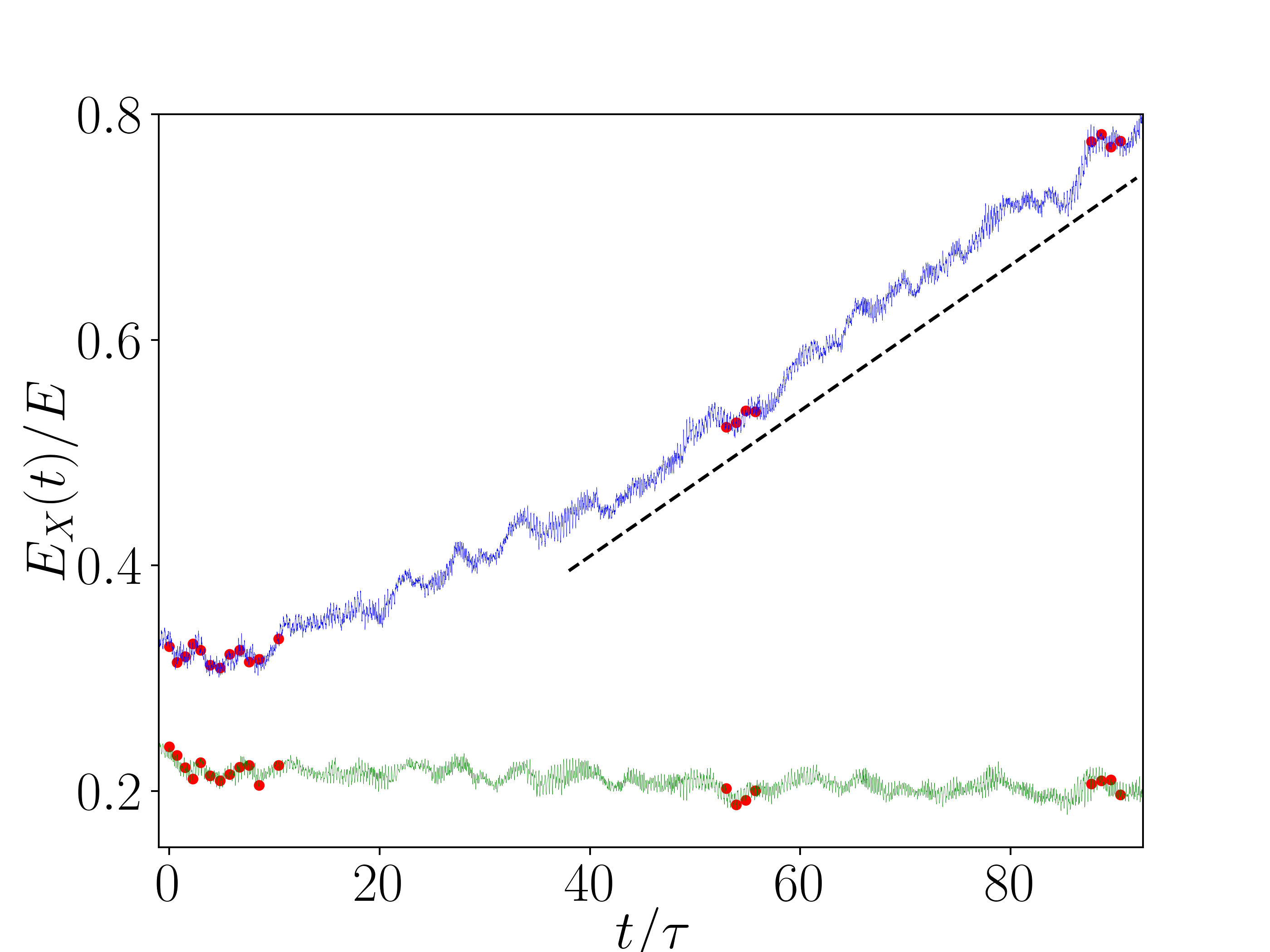}
    \includegraphics[width=0.49\columnwidth]{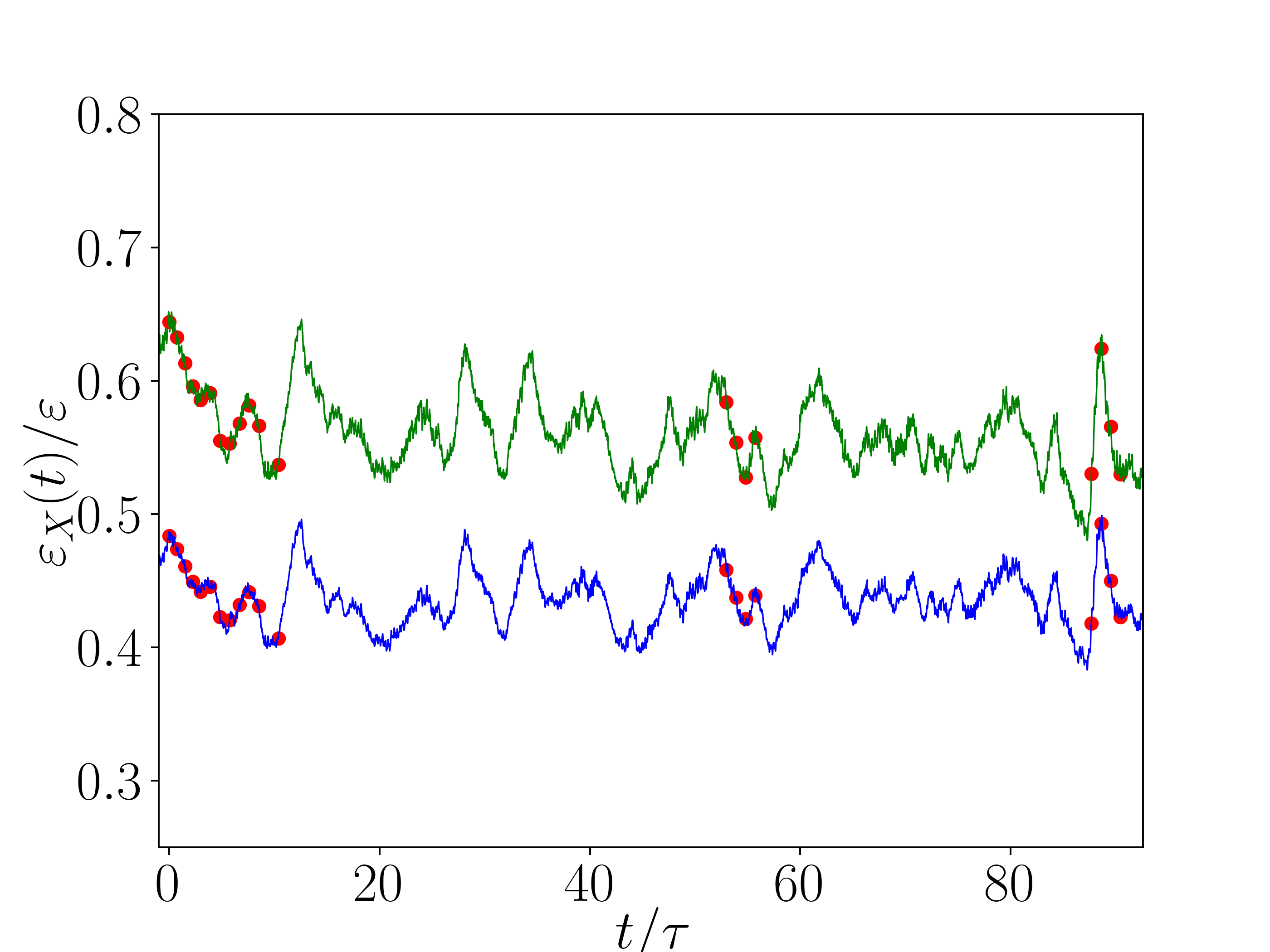}
    \caption{Like fig.~\ref{fig:hyperv_1024}, time evolution of global observables for dataset C10. The black dashed line indicates the linear growth of the mean kinetic energy while the employment of thin lines wants to emphasise the \emph{fast} oscillations of both the magnetic and kinetic energy. Unlike the right panel, the y-axis range of the left panel differs slightly from that of fig.~\ref{fig:hyperv_1024}.}
    \label{fig:B0_10}
\end{figure}

In order to emphasise the effects of the BMF, in fig.~\eqref{fig:visuals} we show the 3D visualisation of the components ${u}_x$ and the magnitude of the electric current $\bm{j} = \nabla \times \bm{b}$ for the datasets A3, C1 and C10. In fact, for the latter we can appreciate a 2Dmensionalization of the
flow by noticing extended coherent
structures along $z-$axis which is the direction of the BMF.  Conversely,
on the plane perpendicular to the $z-$axis, the velocity fluctuations presents patches
that are smaller and more fragmented than those of $B_0=0$ and even of
$B_0=1.2 B_{rms}$. Such an effect is also noticeable in the visualisation of the magnitude of the electric
current $\bm{j}$. On the contrary, the velocity components visualizations related to $B_0 = 1.2\, B_{rms}$ show more \emph{broken} and filamented
patches of current-sheets compared to the $B_0=0$ case where the coherent structures appear to be larger and more monochromatic in the visualisation colouring. The same behaviour can be seen in the corresponding current $\bm{j}$ as well.
\\
Fig.~\ref{fig:spectra_all} shows the time-averaged omnidirectional kinetic and magnetic spectra for each datasets. Here, the employment of hyperviscosity makes the dissipation range more concentrated in the small scales, leading to the sharp fall-off of the spectra.
Although we do not apply a magnetic forcing, the magnetic energy spectra of panels (c) and (d) are characterised by a peak in the forcing wavenumber band because of a spectral coupling due the the non-zero BMF (see e.g. sec. 5.1 of \cite{damiano2024thesis} for further explanation).
Moreover, in the kinetic energy spectrum shown in panel (d), the modes associated with \emph{small} wavenumbers are significantly more excited compared to those in the other configurations. This is due to a partial inverse cascade of kinetic energy, commensurate with the enhanced two-dimensionalization of the fluctuations, as can be expected in presence of a strong BMF. Although the applied forcing scheme differs, the effects of a \emph{strong} BMF have been studied, for instance, by \cite{alexakis2011} and \cite{gallet2015}.
\\
In figs.~\ref{fig:hyperv_1024}-\ref{fig:B0_10} we can observe time series of the total kinetic and magnetic energy per unit volume, $E_u(t) = \frac{1}{2}\langle |\bm{u}(\bm{x},t)|^2 \rangle_V$ and $E_b(t) = \frac{1}{2}\langle |\bm{b}(\bm{x},t)|^2 \rangle_V$, together with the kinetic and magnetic standard energy dissipation rate, $\varepsilon_u(t) = \langle |\nabla \bm{u}(\bm{x},t)|^2 \rangle_V$ and $\varepsilon_b(t) = \ \langle |\nabla \bm{b}(\bm{x},t)|^2 \rangle_V$, for each dataset. The red points correspond to instances in time where full data cubes have been sampled. It is clear from the left panel of fig.~\eqref{fig:B0_10} that the system described by dataset C10 is non-stationary since it exhibits a linear growth in the kinetic energy. Moreover studying the kinetic energy spectrum over time, it is observed that the non-stationarity pertains the low wavenumbers energy modes as the system is accumulating energy in the \emph{large} scales. In this respect, the fields snapshots are sampled in the quasi-stationarity intervals. It is interesting to underline that the \emph{small scale} quantities of the right panel show a stationary behaviour over time. 

\clearpage

\subsection{Reynolds number estimate for hyperviscous runs}

To calculate an effective Reynolds number for the hyperdissipative datasets we follow the methodology described in \cite{BuzzicottiEA18-xfer}. 
There, the standard integral-scale Reynolds number, based on standard Laplacian dissipation, 
  $ Re = {U L_u} / {\nu} \propto (L_u /\eta_1)^{4/3}$
 \citep[e.g.,][]{BatchelorTHT,Pope}
is replaced with one based on the ratio between the integral scale $L_u$ and the effective dissipation range scale $I_d$. In particular, we employ
\begin{equation}\label{eq:michele_reynolds}
   Re = C \left( \dfrac{L_u}{I_d} \right)^{4/3} ,
\end{equation}
where $I_d = \pi/\text{argmax}\left( k^2 {E_u}(k) \right)$ 
is the scale where the dissipation spectrum $k^2 E_u(k) $ 
shows a maximum.
Here, $C$ is a fit parameter that has to be estimated by comparing eq.~\eqref{eq:michele_reynolds} with the common definition of the Reynolds number in a standard-viscosity run. 
According to this procedure we obtain $C=40$ that allows for an estimate of the effective Reynolds number for datasets A3, A4, C1 and C10. In consequence of a different type of flow, it is relevant to underline that a more refined $Re$ estimate for hyper-diffusive MHD configurations with a \emph{strong} BMF would require the calculation of the $C$ constant from the corresponding standard viscosity run. 
\begin{figure}
	\begin{center}
         \noindent\makebox[\textwidth]{
         \hspace{-2.0cm}
         \includegraphics[width=.95\columnwidth]{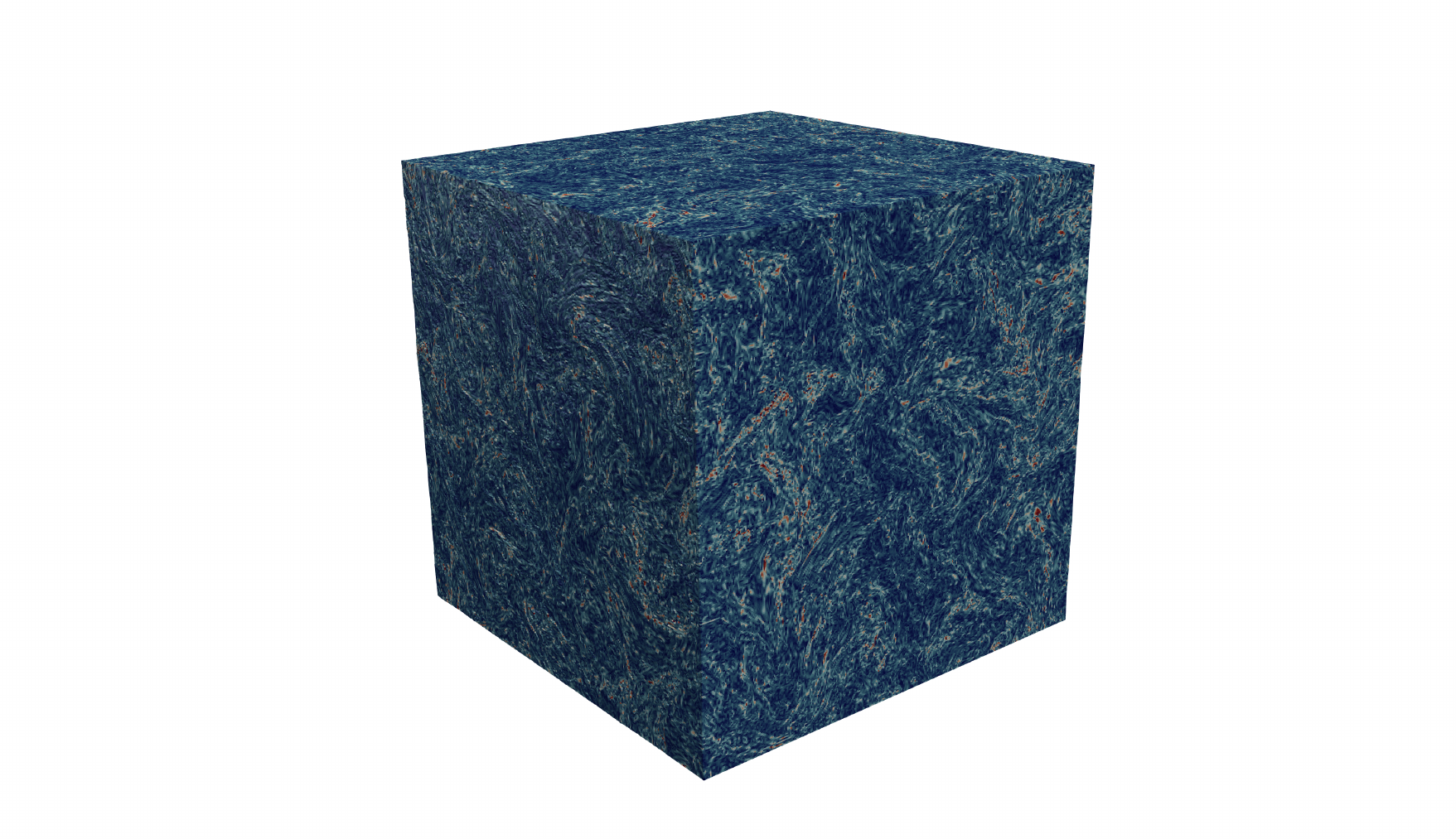}  
         \hspace{-3.6cm}
         \includegraphics[width=.8\columnwidth]{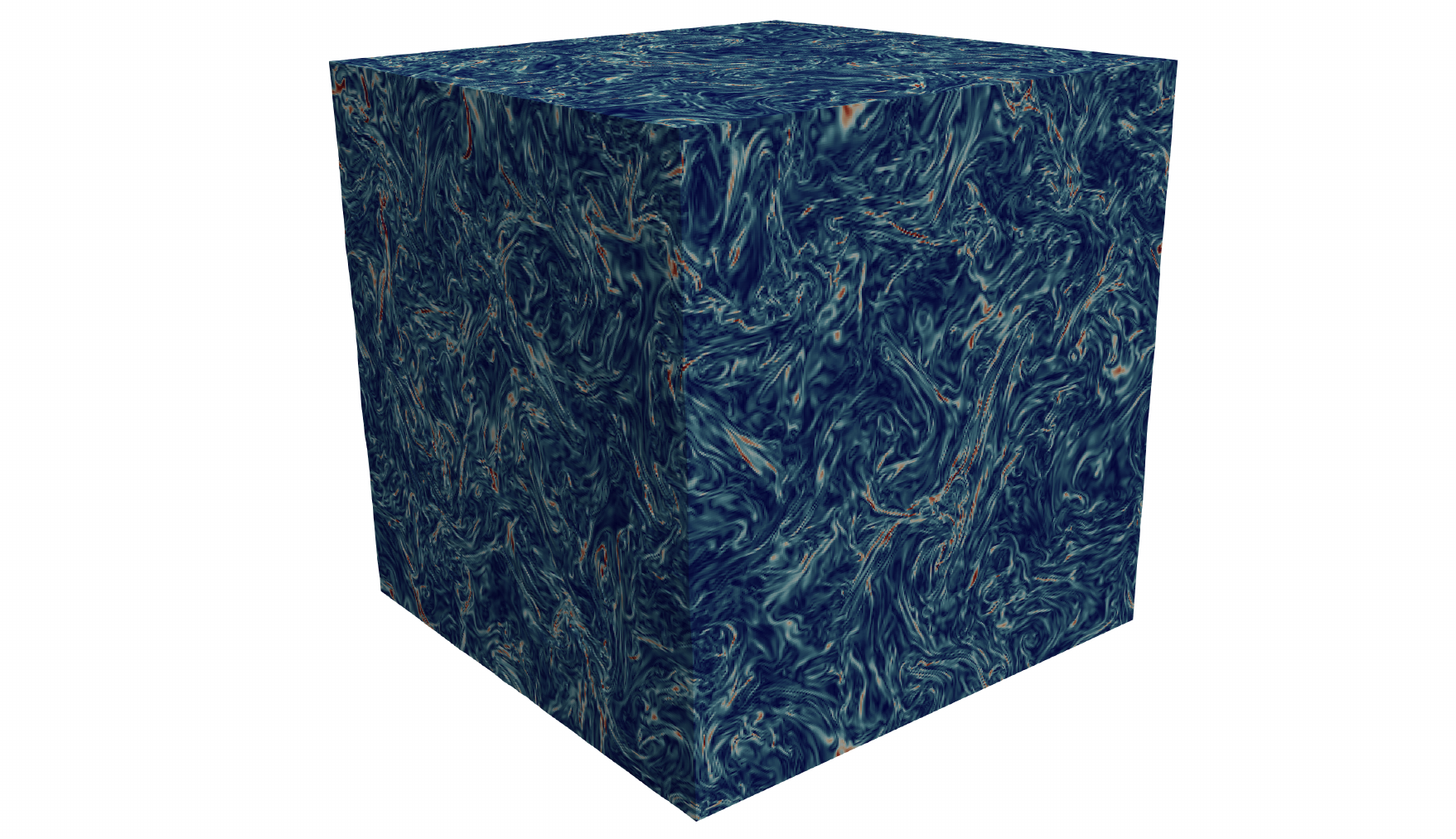} 
         }
    \end{center}
	 \caption{3D visualization of the magnitude of the vorticity $\bm{\omega} = \nabla \times \bm{u}$ as a function of the position $(\bm{x},\bm{y}$, $\bm{z})$ related to one instantaneous field configuration of standard diffusive MHD from dataset A1 (right panel) and hyperdiffusive MHD from dataset A3 (left panel). The two panels share the same colour range.}
\label{fig:visuals_hyp_visc}
\end{figure}
In any case, from the visualisation of one instantaneous velocity configuration in fig.~\ref{fig:visuals_hyp_visc}, we notice that hyperviscosity is associated with a more pronounced separation of scale manifested by the presence of \emph{smaller scale} structures. Furthermore, as a consequence of the $Pm=1$, we observe that the structure formation of the vorticity magnitude, in the left panel of fig.~\ref{fig:visuals_hyp_visc}, is qualitatively similar to that of the electric current in the top-right panel fig.~\ref{fig:visuals}.

\section{DataBase Description}

\subsection{Datasets history}
Datasets A1 and A3 originate from pre-existing MHD datasets with $N=1024^3$ grid-points \citep{sean_priv_comm} employing standard and hyperviscous dissipation laws respectively.\\

\noindent Dataset A4, C1 and C10 originate from the last fields snapshot of A3. In particular, for C1 and C10 the magnitude of the imposed BMF is 1 and 10 respectively.\\

\noindent Dataset A2 used the last snapshots of A1 as initial values.

\subsection{Database files}
TURB-Hel database is made of files extracted from the datasets described in the previous section as follows:

\begin{itemize}
    \item During the simulation we dumped the vector potential of both velocity and magnetic field in Fourier space \textbf{every} simulation steps. The instantaneous configuration have been called "cb" and "ch" (from Complex B-field and Complex H-field). Each dataset is equipped with a text file called selected\_snaps\_file.txt that indicates the simulation time of the snapshots sampling. {Specifically, for datasets A1--A3--C10 time increases with the snapshot index, whereas for A2--A4--C1 it decreases.}

    \item In order to recover both the velocity and magnetic fields, every configuration should be read using the HDF5 library, then a curl operation is needed to compute the fields in Fourier space.

    \item If the fields are needed in real space, a backward FFT is needed.

    \item In the support materials, there is a C program \begin{verbatim} read_cb \end{verbatim} which performs all the above steps, and an accompanying ReadME.pdf for the details on how to compile it. 
\end{itemize}

The database TURB-MHD is available for download using the SMART-Turb portal at \url{http://smart-turb.roma2.infn.it}. 
\\

The SMART-Turb portal also comprises of the following databases: TURB-Hel \citep{turb-hel} describing helically forced homogeneous and isotropic turbulence, TURB-Rot \citep{TURB-Rot} for rotating turbulence and TURB-Lagr \citep{turb-hel} related to Lagrangian particle under turbulence.  \\

\section*{Acknowledgements}
This work received funding from the European Research Council (ERC) under the European Union's Horizon 2020 research and innovation programme (grant agreement No 882340). Computational resources were provided through the UK Turbulence Consortium on ARCHER2 (EPSRC grants EP/R029326/1 and EP/X035484/1). 

\bibliographystyle{jfm}
\bibliography{bib,bib_2,bib_3,bib_4,extra}

\end{document}

%% file: latex-defns.tex
   \newcommand{\vct}[1]  {\ensuremath{\boldsymbol{#1}}}    

 \newcommand{\vb} {\vct{b}}

 \newcommand{\vu} {\vct{u}}
 
 \newcommand{\vx} {\vct{x}}

 \newcommand{\vF} {\vct{F}}

\newcommand{\squishlist}{
   \begin{list}{$\bullet$}
    { \setlength{\itemsep}{0pt}      \setlength{\parsep}{3pt}
      \setlength{\topsep}{3pt}       \setlength{\partopsep}{0pt}
      \setlength{\leftmargin}{1.5em} \setlength{\labelwidth}{1em}
      \setlength{\labelsep}{0.5em} } }

\newcommand{\squishlisttwo}{
   \begin{list}{$\bullet$}
    { \setlength{\itemsep}{0pt}    \setlength{\parsep}{0pt}
      \setlength{\topsep}{0pt}     \setlength{\partopsep}{0pt}
      \setlength{\leftmargin}{2em} \setlength{\labelwidth}{1.5em}
      \setlength{\labelsep}{0.5em} } }

\newcommand{\squishend}{
    \end{list}  }

